\documentclass[conference]{IEEEtran}
\IEEEoverridecommandlockouts
\usepackage{cite}
\usepackage{amsmath,amssymb,amsfonts}
\usepackage{algorithmic}
\usepackage{graphicx}
\usepackage{textcomp}
\usepackage{xcolor}
\usepackage{booktabs}
\usepackage{array}
\usepackage{multirow}
\usepackage[hidelinks]{hyperref}

\def\BibTeX{{\rm B\kern-.05em{\sc i\kern-.025em b}\kern-.08em
    T\kern-.1667em\lower.7ex\hbox{E}\kern-.125emX}}
\begin{document}

\title{When NPUs Are Not Always Faster: A Stage-Level Analysis of Mobile LLM Inference}

\author{Pu Li*, Jiawen Qi*, Qinyu Chen,~\IEEEmembership{Member,~IEEE}
\thanks{*Pu Li and Jiawen Qi equally contribute to the work.}
\thanks{Pu Li, Jiawen Qi and Qinyu Chen are with the Leiden Institute of Advanced Computer Science (LIACS), Leiden University, The Netherlands. (q.chen@liacs.leidenuniv.nl)}
\thanks{The authors acknowledge the Dutch Research Council (NWO) Veni Talent Program (21828) and the LIACS strategic PhD program for partial funding of this work.}
}

\maketitle

\begin{abstract}
Deploying large language models (LLMs) on mobile devices increasingly relies on heterogeneous execution, yet no prior study has systematically characterized NPU effectiveness at the operator and pipeline level.
We present the first stage-aware, multi-level benchmarking study of mobile LLM inference on a CPU--NPU heterogeneous SoC. We introduce OPMASK-based controlled pipeline decomposition methodology that isolates communication, quantization, and computation overheads within the NPU execution path.
Our results reveal a counter-intuitive stage-level performance reversal: CPUs outperform NPUs in the compute-intensive Prefill stage (up to 1.6x), while NPUs provide only limited acceleration in the memory-bound Decode stage (1.05--1.2x). We further show that scheduling overhead and cross-backend fallback reduce the practical benefits of NPU offloading.
For the energy trend, increasing NPU offloading leads to higher energy consumption (up to 51\%).
Based on these findings, we derive design guidelines for NPU architects targeting on-device LLM inference. 
\end{abstract}


\section{Introduction}
Large language models (LLMs) have recently achieved remarkable success in natural language processing~\cite{S1}, and are increasingly being deployed beyond cloud to edge and mobile devices~\cite{S2,M1,M4}. On-device LLM inference reduces latency, enhances privacy, and eliminates dependence on network connectivity. However, mobile platforms are constrained by limited compute capability, memory bandwidth, and power budget, making LLM deployment a significant challenge.

Existing research on mobile LLM inference can be broadly categorized into three directions. First, model-level optimization techniques, such as quantization and compression, reduce computational and memory overhead~\cite{Q1,Q2,Q3,Q4}. Second, system-level optimization improves efficiency through better scheduling, memory management, and hardware-aware design~\cite{E1,E3,E4,E6,H1}. Third, hardware-level benchmarking evaluates performance across devices and configurations~\cite{B1,B3,E8}. 
Our work falls into the third direction. 
Table~\ref{tab:related} positions our work relative to prior mobile LLM benchmarking studies across five analysis dimensions.
Prior studies cover only parts of this design space. MLPerf Mobile~\cite{B1} reports aggregate throughput only; Zhang and Huang~\cite{B2} compare CPU and GPU but not NPUs; Chen et al.~\cite{B3}, llm.npu~\cite{E4}, and Hao et al.~\cite{E8} study cooperative or pure-NPU systems without operator-level competitive CPU--NPU profiling. No prior work combines competitive CPU--NPU comparison, stage awareness, operator-level profiling, and pipeline decomposition.

To address these gaps, we study mobile LLM inference on a heterogeneous CPU--NPU platform, since NPU behavior under LLM workloads remains poorly characterized despite its growing importance in on-device AI.
This paper makes the following contributions:
\begin{itemize}
\item We present the first stage-aware benchmarking study of mobile LLM inference on CPU--NPU heterogeneous platforms, covering four models across two architecture families with systematic variation of offloading ratio and prompt length.

\item We introduce the OPMASK-based pipeline decomposition methodology that isolates communication, quantization, and computation overheads within NPU execution.

\item We quantify two previously uncharacterized bottlenecks (1) scheduling tax (call-usec 8--22$\times$ of op-usec for lightweight operators) and (2) cross-backend fallback penalty (1.5$\times$) that erode the NPU's core advantage.

\item We derive three design guidelines for NPU architects and system integrators, grounded in quantitative evidence from our measurements.
\end{itemize}

\begin{table}[t]
\centering
\setlength{\tabcolsep}{1.5pt}
\caption{Comparison with prior mobile LLM benchmarking studies. SA: stage-aware; OP: operator-level profiling; PD: pipeline decomposition; EN: energy analysis.}
\label{tab:related}
\begin{tabular}{l l l c c c c}
\toprule
\textbf{Study} & \textbf{Backends} & \textbf{Type} & \textbf{SA} & \textbf{OP} & \textbf{PD} & \textbf{EN} \\
\midrule
MLPerf Mobile~\cite{B1}      & GPU, NPU        & Independent   & -- & -- & -- & -- \\
Zhang \& Huang~\cite{B2}     & CPU, GPU        & Competitive   & -- & -- & -- & -- \\
Chen et al.~\cite{B3}        & GPU+NPU         & Cooperative   & \checkmark & -- & -- & \checkmark \\
llm.npu~\cite{E4}            & CPU+GPU+NPU     & Cooperative   & -- & -- & -- & -- \\
Hao et al.~\cite{E8}         & NPU             & Single        & -- & -- & -- & -- \\
\textbf{This work}           & \textbf{CPU, NPU} & \textbf{Competitive} & \checkmark & \checkmark & \checkmark & \checkmark \\
\bottomrule
\end{tabular}
\end{table}

\section{Methodology}

\begin{figure}[t]
\centering
\includegraphics[width=0.8\columnwidth]{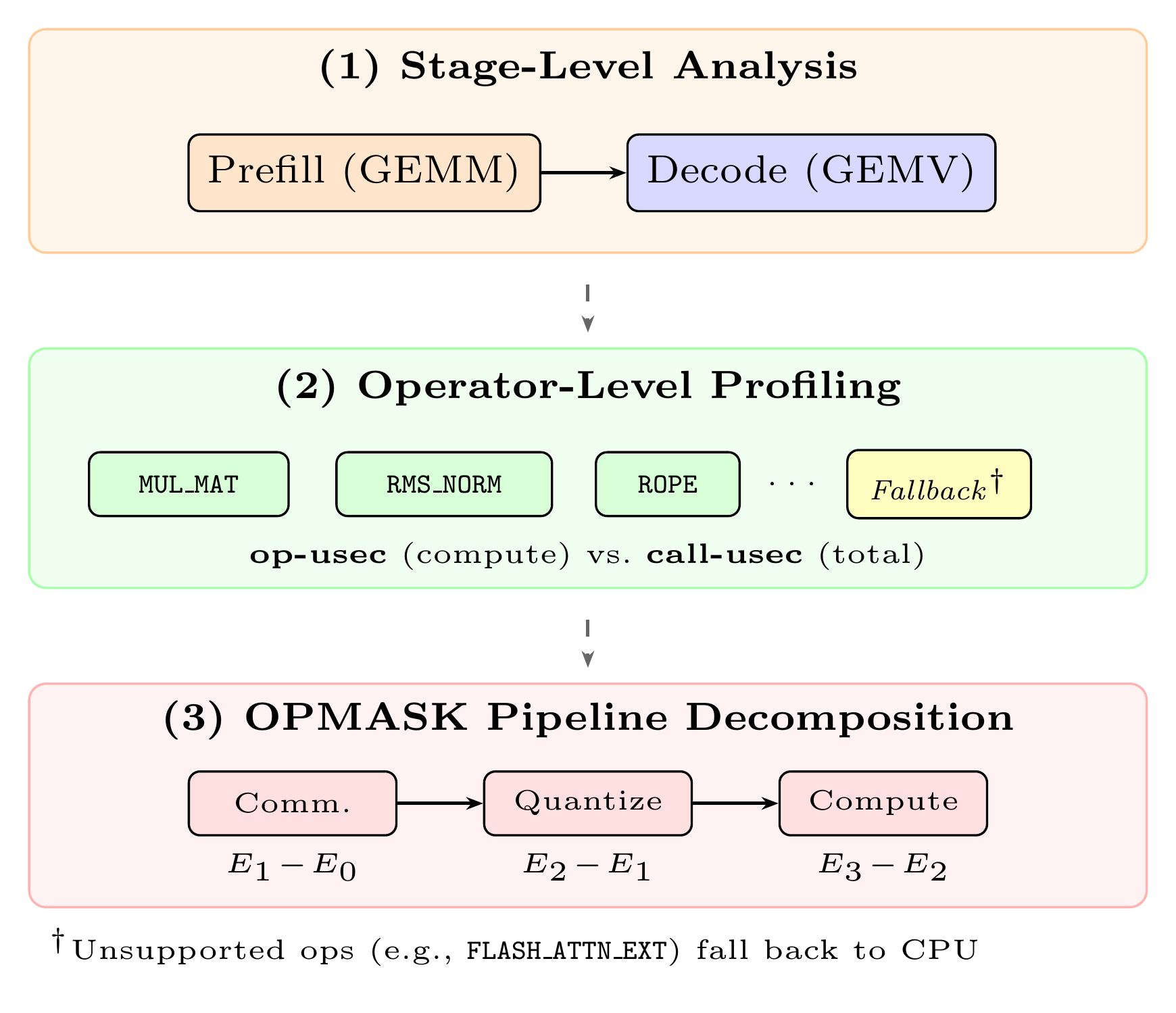}
\caption{Overview of the multi-level analysis framework. Stage-level analysis separates Prefill and Decode; operator-level profiling compares compute time vs.\ scheduling overhead; OPMASK decomposes the NPU pipeline via controlled masking.}
\label{fig:overview}
\end{figure}

Fig.~\ref{fig:overview} presents the overview of our multi-level analysis framework.
Our goal is not only to compare CPU and NPU throughput at the system level, but also to explain the source of their performance differences.

\subsection{System Overview}
We build our analysis framework on top of the \texttt{llama.cpp}~\cite{E2} inference engine, which represents the model forward pass as a static computation graph. Each node corresponds to an operator, and execution is scheduled across heterogeneous backends.
During execution, a backend scheduler partitions the computation graph based on operator support. Supported operators are dispatched to the NPU, while unsupported ones fall back to the CPU, resulting in a heterogeneous execution pipeline. This execution model introduces additional overhead due to graph partitioning, synchronization, and data transfer.

\subsection{Stage-Level Execution Model}
LLM inference consists of two distinct stages:
(1) \textbf{Prefill stage}: processes the entire input sequence using batched matrix--matrix multiplication (GEMM), exhibiting high compute intensity and parallelism.
(2) \textbf{Decode stage}: generates tokens autoregressively using matrix--vector multiplication (GEMV), characterized by strong sequential dependency and memory-bound behavior.
We model the total inference latency as
$T_{\text{total}} = T_{\text{prefill}} + T_{\text{decode}}$
This distinction is critical, as the two stages exhibit fundamentally different computational characteristics.

\subsection{Operator-Level Profiling}
To further attribute performance differences, we introduce operator-level profiling that captures both computation time and system-level overhead. Operators include MUL\_MAT, RMS\_NORM, ROPE, etc.
For each operator, we record (1) op-usec, representing the actual execution time on the compute unit;
(2) call-usec, representing the total invocation time, including scheduling, communication, and synchronization overhead.

By comparing these two metrics, we quantify the gap between pure computation and end-to-end execution. In particular, a large call-usec / op-usec ratio indicates that performance is dominated by orchestration overhead rather than arithmetic efficiency.
This distinction is critical for heterogeneous execution, where frequent CPU--NPU interactions can introduce significant overhead even when individual operators are fast.

\subsection{OPMASK-Based Pipeline Decomposition}
While operator-level profiling reveals per-operator overhead, it does not directly explain how latency is distributed within the NPU execution pipeline. To address this, we introduce an OPMASK-based decomposition method.
We model each NPU operator execution as a three-step pipeline:
(1) communication, corresponding to CPU--NPU request dispatch and response handling;
(2) quantization, corresponding to dynamic data format conversion (e.g., F32 $\rightarrow$ INT8);
(3) computation, corresponding to execution on NPU compute units.
We use four execution modes (see Table~\ref{tab:opmask_conditions}) with progressively enabled functionality to calculate the latency of each component:
$T_{\text{comm}} = E_1 - E_0,
T_{\text{quant}} = E_2 - E_1,
T_{\text{compute}} = E_3 - E_2.$
This approach decomposes tightly coupled execution stages without modifying hardware, providing a quantitative view of where time is spent within the NPU pipeline.


\begin{table}[t]
\centering
\renewcommand{\arraystretch}{0.8}
\setlength{\tabcolsep}{2pt}
\caption{OPMASK Experimental Conditions}
\label{tab:opmask_conditions}
\begin{tabular}{l c l p{3.0cm}}
\toprule
\textbf{Cond.} & \textbf{OPMASK} & \textbf{Stages} & \textbf{Observation Target} \\
\midrule
E0 (baseline) & 0x0 & None & CPU-side graph traversal and scheduling overhead \\
E1 & 0x1 & Queue & CPU--NPU communication round-trip overhead \\
E2 & 0x3 & Queue + Quantize & Comm. + dynamic quantization cumulative overhead \\
E3 (default) & 0x7 & All & Full NPU execution time \\
\bottomrule
\end{tabular}
\end{table}

\section{Experimental Setup}

All experiments are conducted on a smartphone equipped with the Snapdragon 8 Gen~3 SoC (1 Prime + 5 Performance + 2 Efficiency CPU cores), Hexagon v75 NPU (HTP architecture with HVX and HMX units), 16\,GB RAM, and Android~15. We evaluate four Q4\_0-quantized models: Llama-3.2-3B (27 repeating layers)~\cite{M1}, Llama-3.1-8B (31 repeating layers), Qwen3-4B (35 repeating layers), and Qwen3-8B (35 repeating layers)~\cite{M4}, using llama.cpp~\cite{E2} (tag: b7588).

Performance is measured separately for the Prefill and Decode stages, with throughput (tokens/s) as the primary metric. Each configuration is executed 10--15 times and results are reported as arithmetic means after IQR-based outlier filtering. The NPU offloading ratio is controlled via the \texttt{ngl} parameter, specifying the number of Transformer layers offloaded to the NPU. CPU threads are set to 6, batch size to 128, and prompt length to $\sim$1\,281 tokens unless otherwise stated. Profiling experiments use synchronous execution mode (\texttt{GGML\_HEXAGON\_OPSYNC=1}) to obtain accurate per-operator timing.

\section{Results and Analysis}

\subsection{System-Level Throughput}

\begin{figure*}[t]
\centering
\begin{minipage}[t]{0.24\textwidth}\centering
\includegraphics[width=\textwidth]{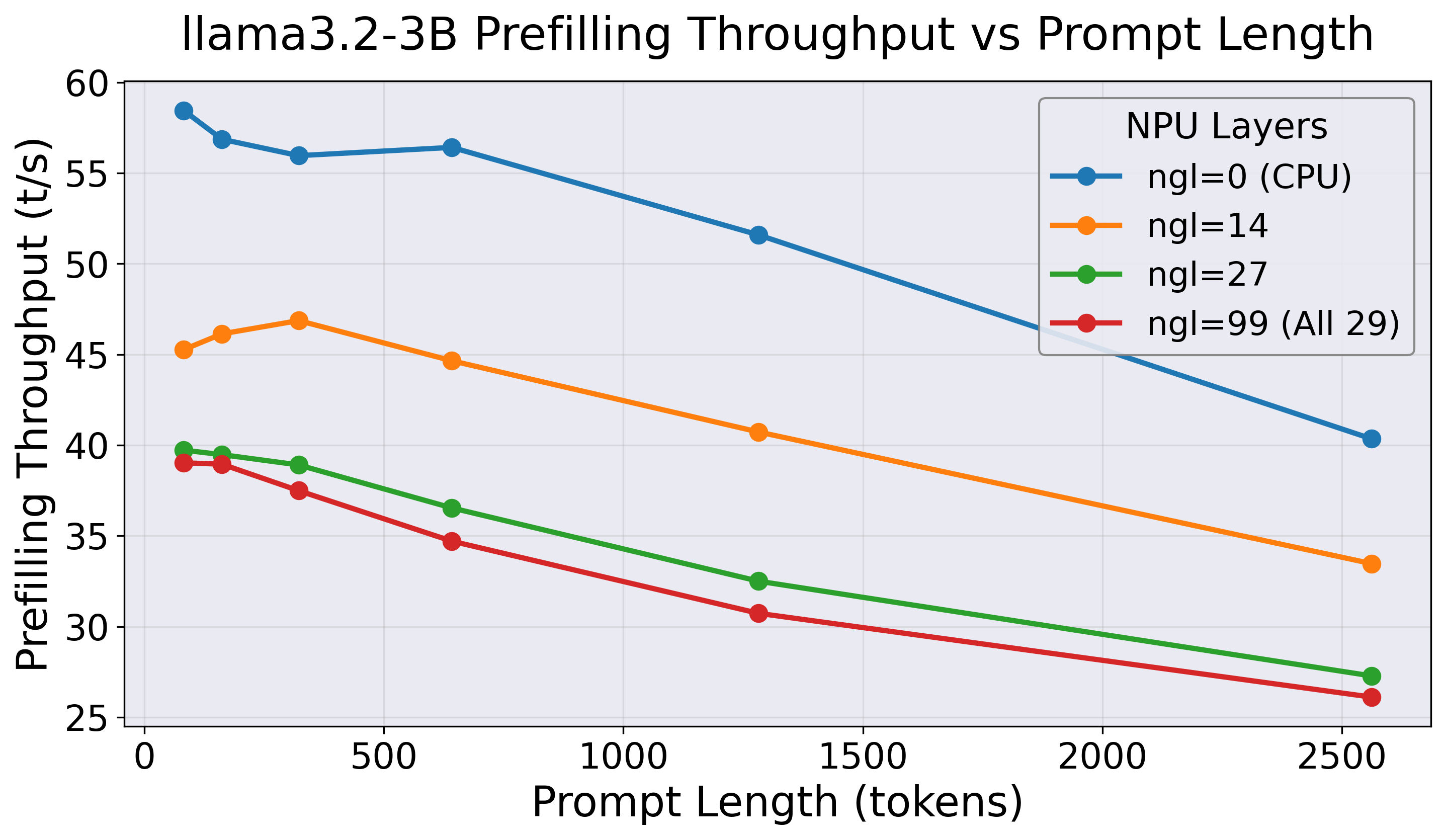}
\centerline{\scriptsize (a) Llama-3.2-3B Prefill}
\end{minipage}\hfill
\begin{minipage}[t]{0.24\textwidth}\centering
\includegraphics[width=\textwidth]{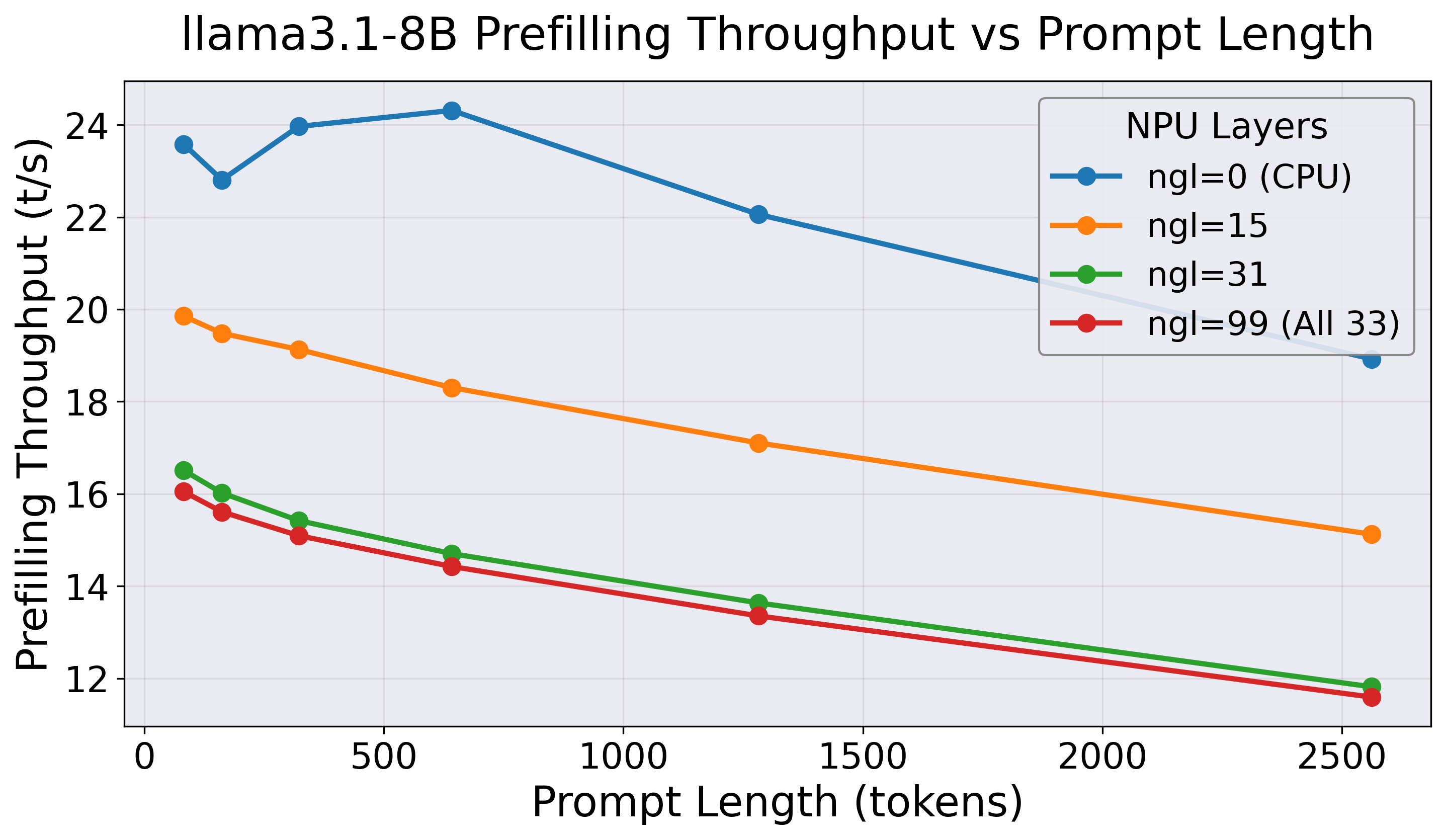}
\centerline{\scriptsize (b) Llama-3.1-8B Prefill}
\end{minipage}\hfill
\begin{minipage}[t]{0.24\textwidth}\centering
\includegraphics[width=\textwidth]{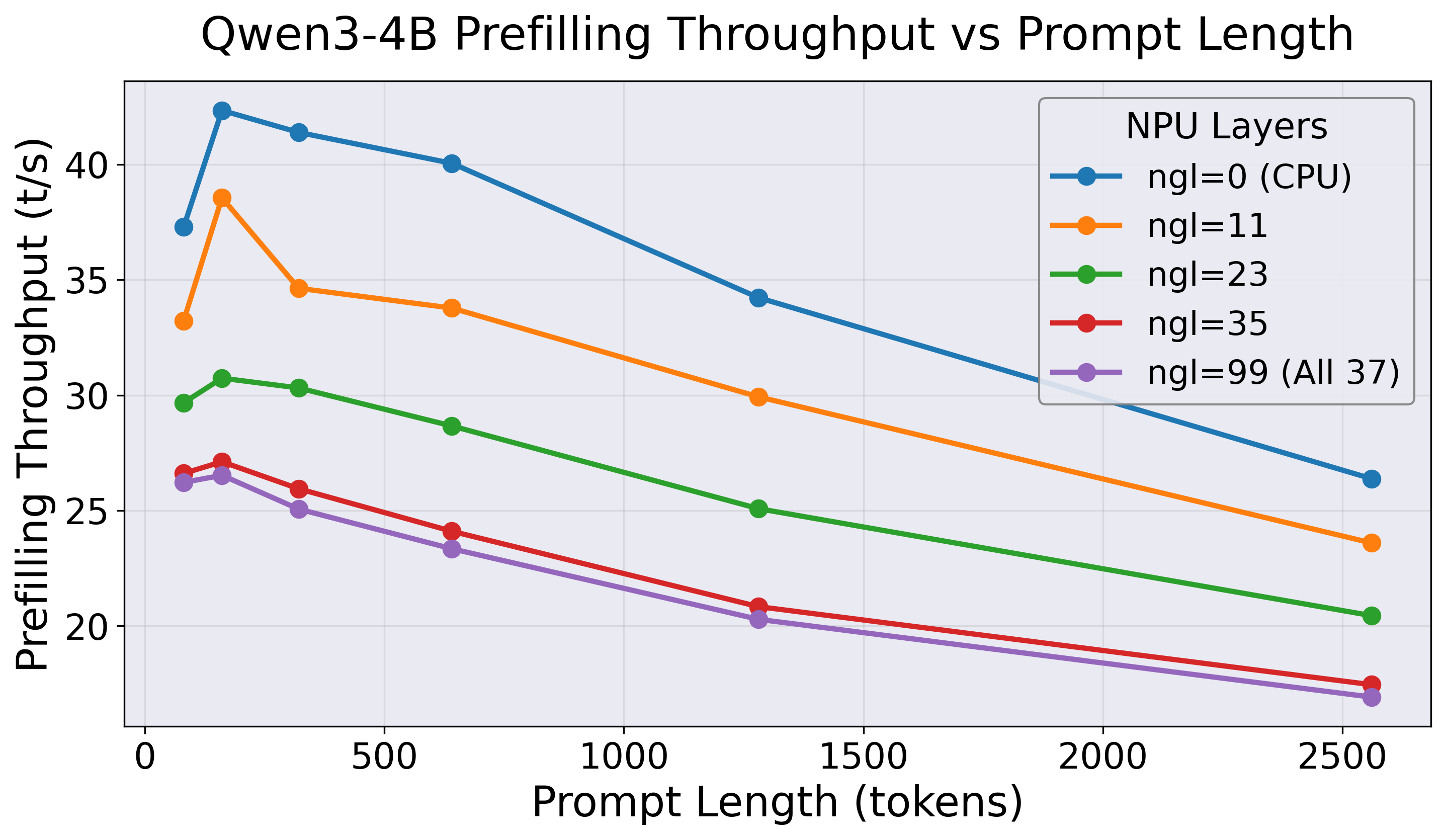}
\centerline{\scriptsize (c) Qwen3-4B Prefill}
\end{minipage}\hfill
\begin{minipage}[t]{0.24\textwidth}\centering
\includegraphics[width=\textwidth]{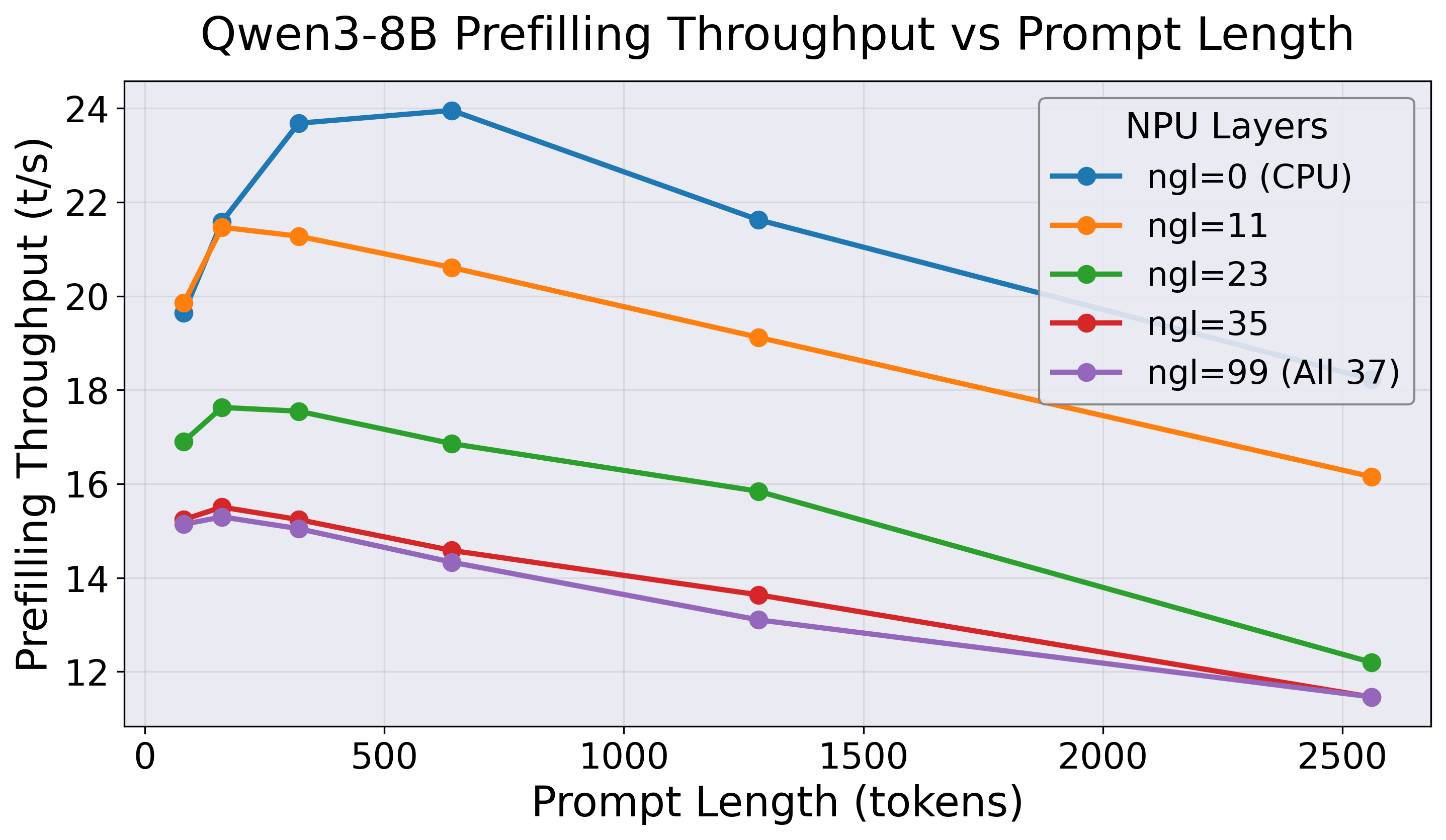}
\centerline{\scriptsize (d) Qwen3-8B Prefill}
\end{minipage}

\vspace{0.1cm}

\begin{minipage}[t]{0.24\textwidth}\centering
\includegraphics[width=\textwidth]{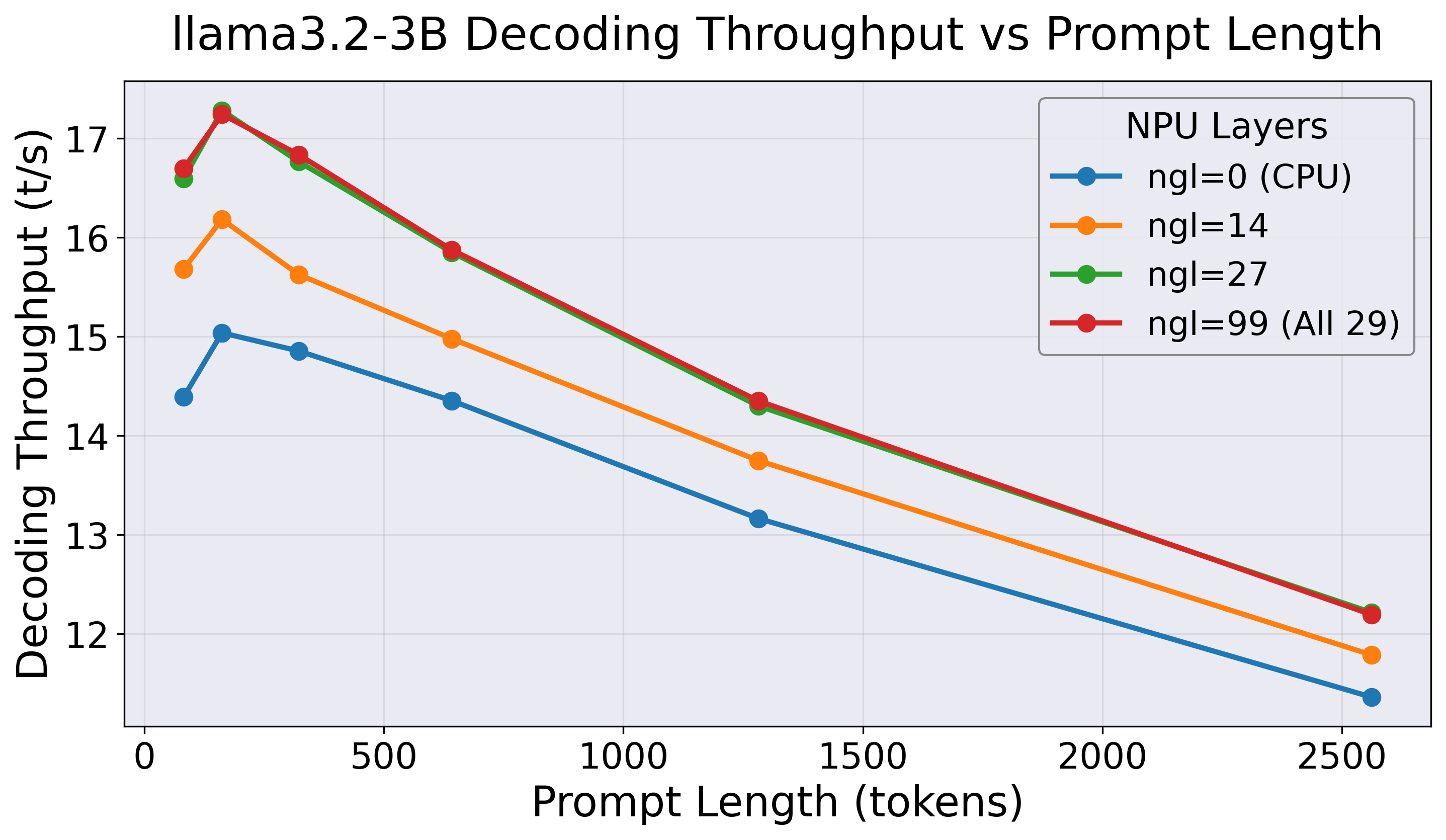}
\centerline{\scriptsize (e) Llama-3.2-3B Decode}
\end{minipage}\hfill
\begin{minipage}[t]{0.24\textwidth}\centering
\includegraphics[width=\textwidth]{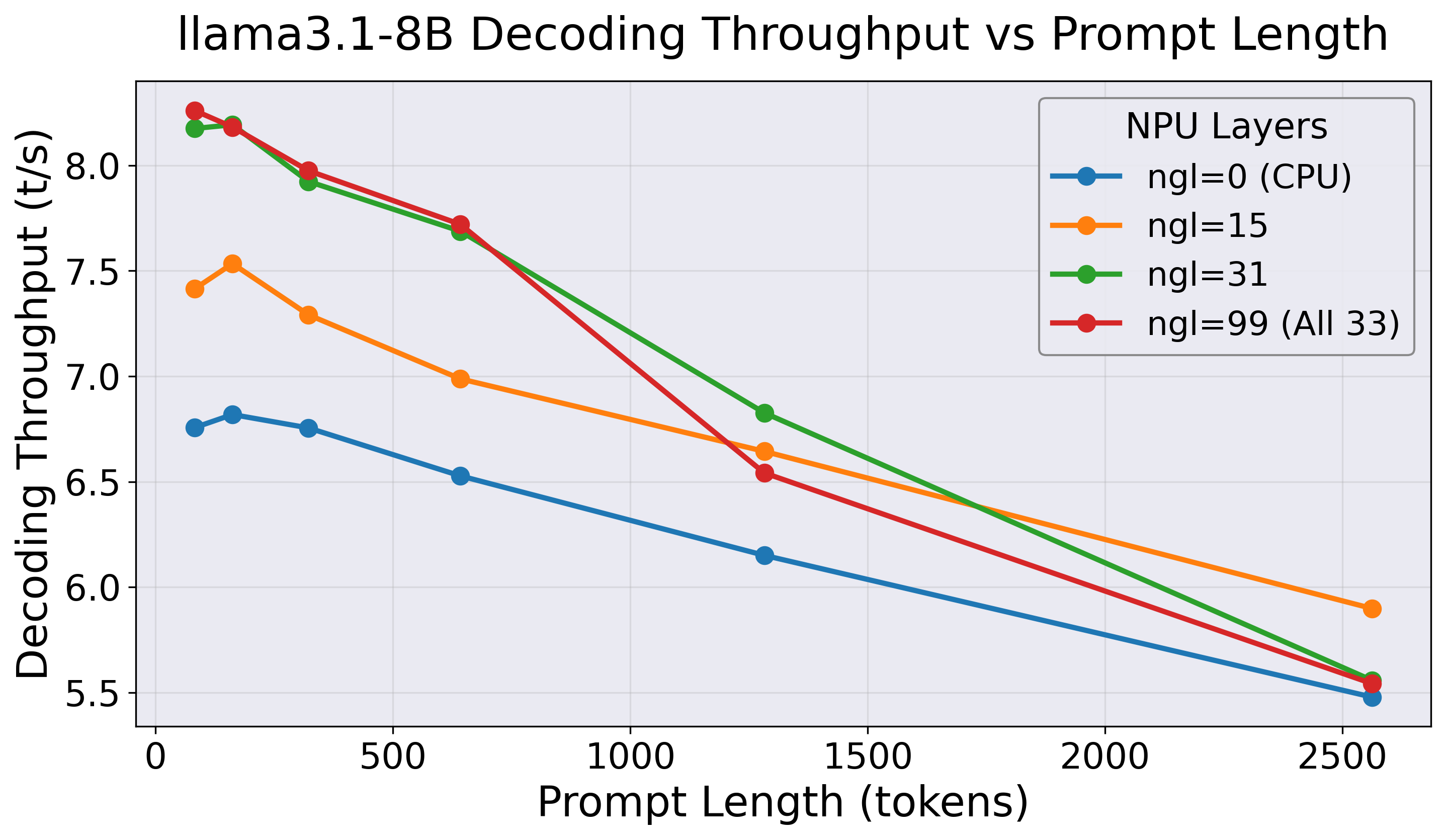}
\centerline{\scriptsize (f) Llama-3.1-8B Decode}
\end{minipage}\hfill
\begin{minipage}[t]{0.24\textwidth}\centering
\includegraphics[width=\textwidth]{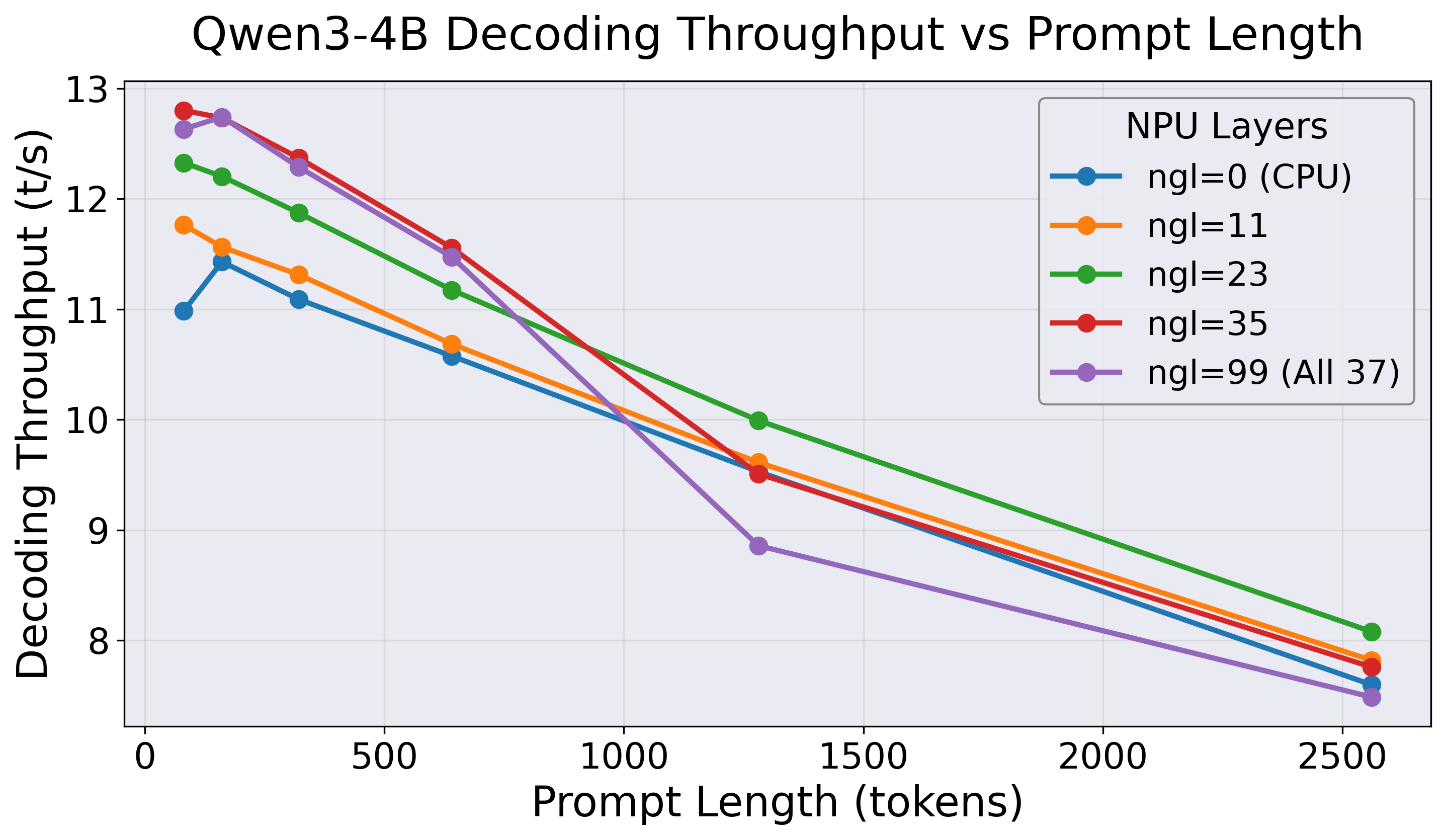}
\centerline{\scriptsize (g) Qwen3-4B Decode}
\end{minipage}\hfill
\begin{minipage}[t]{0.24\textwidth}\centering
\includegraphics[width=\textwidth]{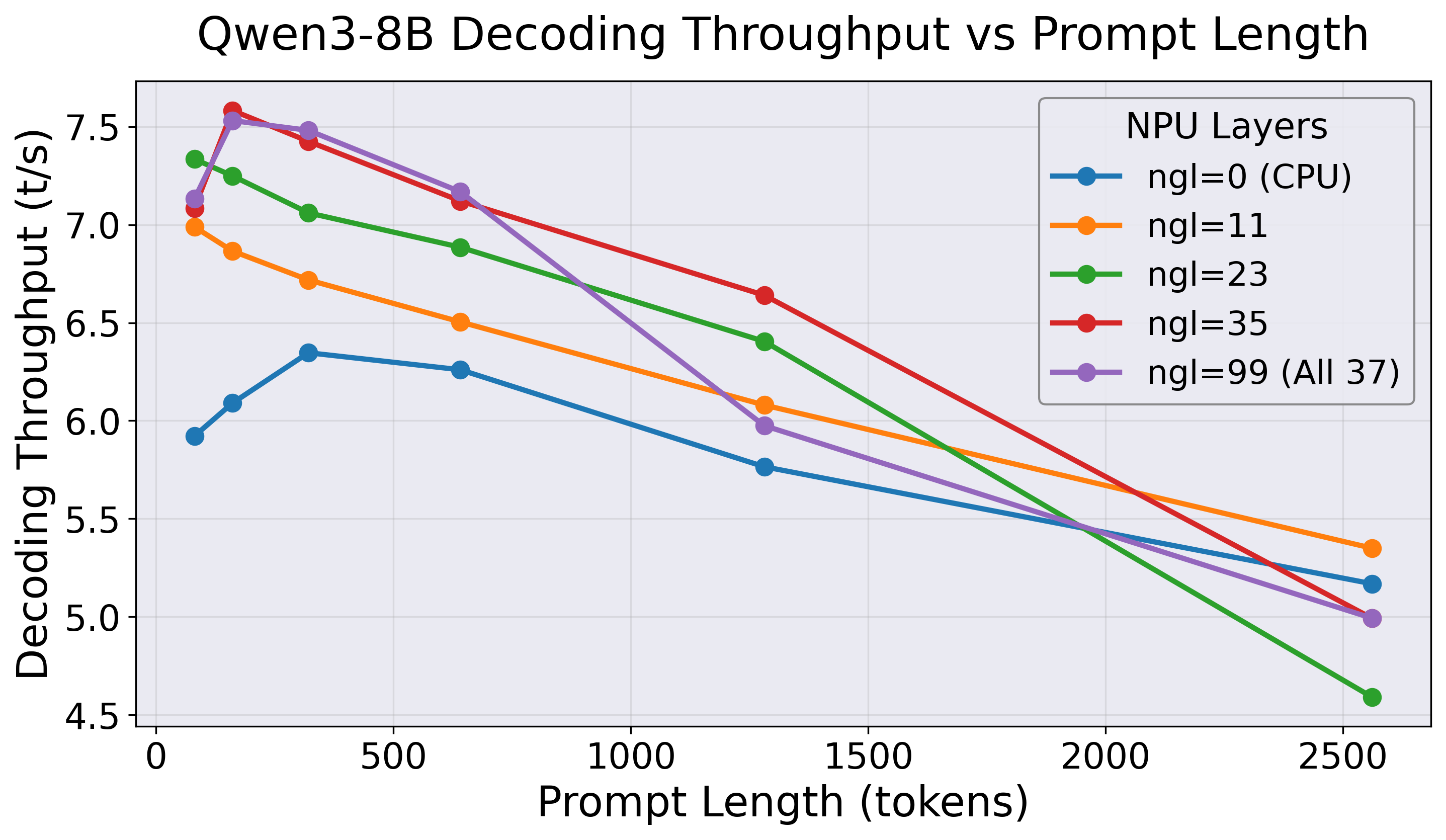}
\centerline{\scriptsize (h) Qwen3-8B Decode}
\end{minipage}

\vspace{0.15cm}


\caption{System-level throughput under varying NPU offloading ratios (\texttt{ngl}) and prompt lengths. Top row: Prefill; bottom row: Decode. CPU-only execution maximizes Prefill throughput, while higher \texttt{ngl} improves Decode throughput.}
\label{fig:throughput}
\end{figure*}

Fig.~\ref{fig:throughput} presents the system-level throughput of all four models under varying NPU offloading configurations (\texttt{ngl}) and prompt lengths. Two contrasting patterns emerge consistently across all models. In the Prefill stage (top row), CPU-only execution (\texttt{ngl=0}) achieves the highest throughput across all prompt lengths, with performance degrading as more layers are offloaded to the NPU. In contrast, the Decode stage (bottom row) shows the opposite: higher \texttt{ngl} values improve throughput, confirming that NPU offloading benefits autoregressive generation. The 3B/4B models achieve higher Prefill throughput than the 8B models, while 8B models show less pronounced degradation under long-sequence conditions. This consistent stage-level divergence motivates our operator-level investigation below.

\subsection{Core Operator Performance Reversal}

Our operator-level profiling reveals a counter-intuitive performance reversal on the dominant \texttt{MUL\_MAT} operator. Table~\ref{tab:mulmat_combined} presents the per-invocation latency across both stages and all four models.

\begin{table}[t]
\centering
\renewcommand{\arraystretch}{0.6}
\setlength{\tabcolsep}{1.5pt}
\caption{Per-invocation \texttt{MUL\_MAT} latency: CPU vs.\ NPU. NPU latency uses call-usec (includes communication overhead).}
\label{tab:mulmat_combined}
\begin{tabular}{l r r c r r c}
\toprule
& \multicolumn{3}{c}{\textbf{Prefill} ($\mu$s)} & \multicolumn{3}{c}{\textbf{Decode} ($\mu$s)} \\
\cmidrule(lr){2-4} \cmidrule(lr){5-7}
\textbf{Model} & CPU & NPU & NPU/CPU Ratio & CPU & NPU & CPU/NPU Ratio \\
\midrule
Llama-3.2-3B & 5,896 & 9,541 & 1.62$\times$ & 348 & 208 & 1.67$\times$ \\
Llama-3.1-8B & 16,085 & 20,378 & 1.27$\times$ & 646 & 404 & 1.60$\times$ \\
Qwen3-4B & 7,181 & 10,352 & 1.44$\times$ & 324 & 210 & 1.55$\times$ \\
Qwen3-8B & 12,033 & 18,444 & 1.53$\times$ & 589 & 359 & 1.64$\times$ \\
\bottomrule
\end{tabular}
\end{table}

In the \textbf{Prefill stage}, where the core workload is batched GEMM, the NPU is consistently 1.27$\times$--1.62$\times$ slower than the 6-core CPU. Three factors contribute to this disadvantage: (1)~the Hexagon backend's HVX Intrinsics implementation has not undergone the extensive community optimization of the CPU path; (2)~the limited on-chip tightly coupled memory (8\,MiB VTCM) constrains data tiling efficiency for the large matrices characteristic of Prefill; and (3)~unsupported operators (e.g., \texttt{FLASH\_ATTN\_EXT}) fall back to the CPU with additional overhead due to cross-backend cache synchronization and tensor reordering.

In the \textbf{Decode stage}, the workload degenerates to matrix--vector multiplication (GEMV). The NPU achieves 1.55$\times$--1.67$\times$ lower latency on the core \texttt{MUL\_MAT} operator. This advantage is consistent with fundamental differences in memory hierarchy design. Decode-stage weights exhibit a pure streaming access pattern, each element is read once and never reused, which is unfavorable for the CPU's hardware-managed cache hierarchy: weights far exceed the 12\,MB L3 capacity, making cache misses and coherency overhead likely. In contrast, the Hexagon NPU uses software-managed scratchpad memory (VTCM) with a dedicated DMA engine that performs bulk transfers using double-buffering, which likely reduces cache-miss-related penalties and improves effective DRAM bandwidth utilization for this workload~\cite{B3}. In the Prefill stage, this advantage disappears because the large input batch increases arithmetic intensity and enables weight reuse across tokens, making the CPU cache hierarchy beneficial rather than detrimental. Despite the NPU's core operator advantage in Decode, it does not translate into proportional end-to-end gains, as analyzed in Sections~\ref{subsec:pipeline} and~\ref{subsec:scheduling}.

\subsection{NPU Pipeline Decomposition}
\label{subsec:pipeline}

We decompose the NPU execution pipeline into communication, quantization, and computation. Table~\ref{tab:opmask_ratio} presents the overhead ratio breakdown across all four models for both Prefill and Decode stages. Core computation dominates, accounting for 85\%--99\% of total NPU execution time. Dynamic quantization overhead is negligible (0.4\%--2.5\%), confirming that the current framework's quantization strategy introduces no significant penalty.
Critically, the communication overhead exhibits a pronounced \textbf{stage-specific asymmetry}. In the Prefill stage, the massive computation volume masks communication latency, which accounts for only $\sim$0.2\%--3.2\% of total time. In the Decode stage, however, the high-frequency dispatch of fine-grained matrix--vector operations causes severe boundary ``communication congestion,'' with communication overhead rising to 9.9\%--13.0\%. 
\begin{table}[t]
\centering
\renewcommand{\arraystretch}{0.5}
\setlength{\tabcolsep}{2pt}
\caption{OPMASK pipeline overhead ratio (\%) for Prefill (P) and Decode (D) stages. Communication overhead surges in Decode.}
\label{tab:opmask_ratio}
\begin{tabular}{l c r r r}
\toprule
\textbf{Model} & \textbf{Stage} & \textbf{Comm.\%} & \textbf{Quant.\%} & \textbf{Compute\%} \\
\midrule
\multirow{2}{*}{Llama-3.2-3B} & P & 0.8 & 0.9 & \textbf{98.3} \\
                               & D & 13.0 & 1.0 & \textbf{86.0} \\
\midrule
\multirow{2}{*}{Llama-3.1-8B} & P & 0.5 & 2.1 & \textbf{97.4} \\
                               & D & 10.5 & 2.5 & \textbf{87.0} \\
\midrule
\multirow{2}{*}{Qwen3-4B}     & P & 0.2 & 0.4 & \textbf{99.4} \\
                               & D & 12.7 & 2.2 & \textbf{85.1} \\
\midrule
\multirow{2}{*}{Qwen3-8B}     & P & 3.2 & $\approx$0 & \textbf{98.5} \\
                               & D & 9.9 & 0.5 & \textbf{89.5} \\
\bottomrule
\end{tabular}
\end{table}


\subsection{Scheduling Overhead and End-to-End Impact}
\label{subsec:scheduling}

\begin{table}[t]
\centering
\setlength{\tabcolsep}{2pt}
\caption{Decode-stage aggregated operator latency.}
\label{tab:decode_e2e}
\begin{tabular}{l c c c c c}
\toprule
& \textbf{CPU} & \multicolumn{3}{c}{\textbf{NPU path breakdown (ms)}} & \\
\cmidrule(lr){3-5}
\textbf{Model} & \textbf{path (ms)} & NPU ops & Fallback to CPU & Sum & \textbf{Ratio} \\
\midrule
Llama-3.2-3B & 19,597 & 11,844 & 5,360 & 17,204 & 1.14$\times$ \\
Llama-3.1-8B & 40,011 & 24,583 & 8,858 & 33,441 & 1.20$\times$ \\
Qwen3-4B     & 23,889 & 15,767 & 6,966 & 22,733 & 1.05$\times$ \\
Qwen3-8B     & 41,285 & 25,827 & 9,839 & 35,666 & 1.16$\times$ \\
\bottomrule
\end{tabular}
\end{table}

Despite the NPU's 1.55--1.67$\times$ advantage on the core \texttt{MUL\_MAT} operator in Decode, the actual end-to-end speedup is only 1.05$\times$--1.20$\times$ (see Table~\ref{tab:decode_e2e}). Two mechanisms explain this gap.

\begin{figure}[t]
\centering
\begin{minipage}[t]{0.48\linewidth}\centering
\includegraphics[width=\linewidth]{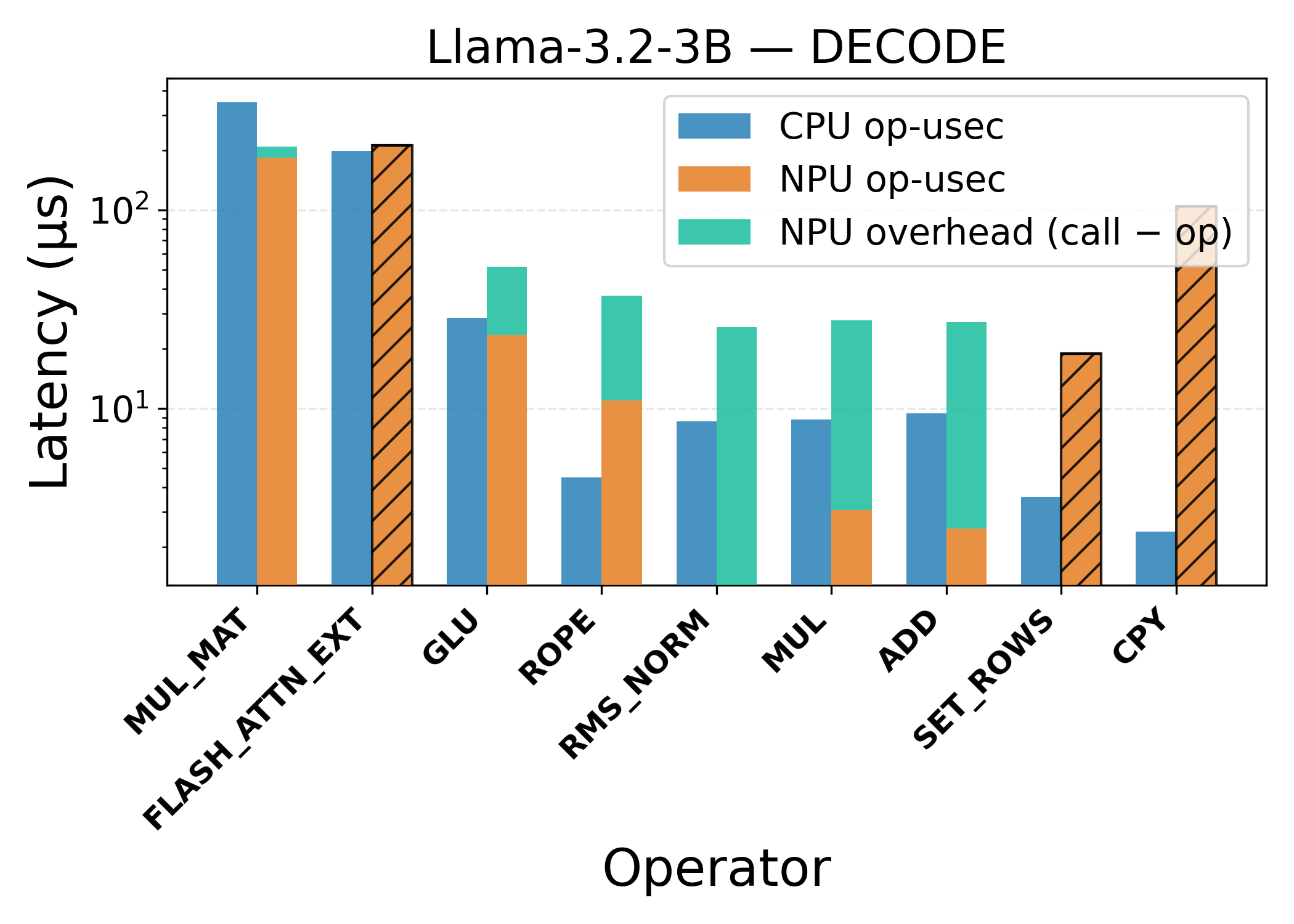}
\centerline{\scriptsize (a) Llama-3.2-3B}
\end{minipage}\hfill
\begin{minipage}[t]{0.48\linewidth}\centering
\includegraphics[width=\linewidth]{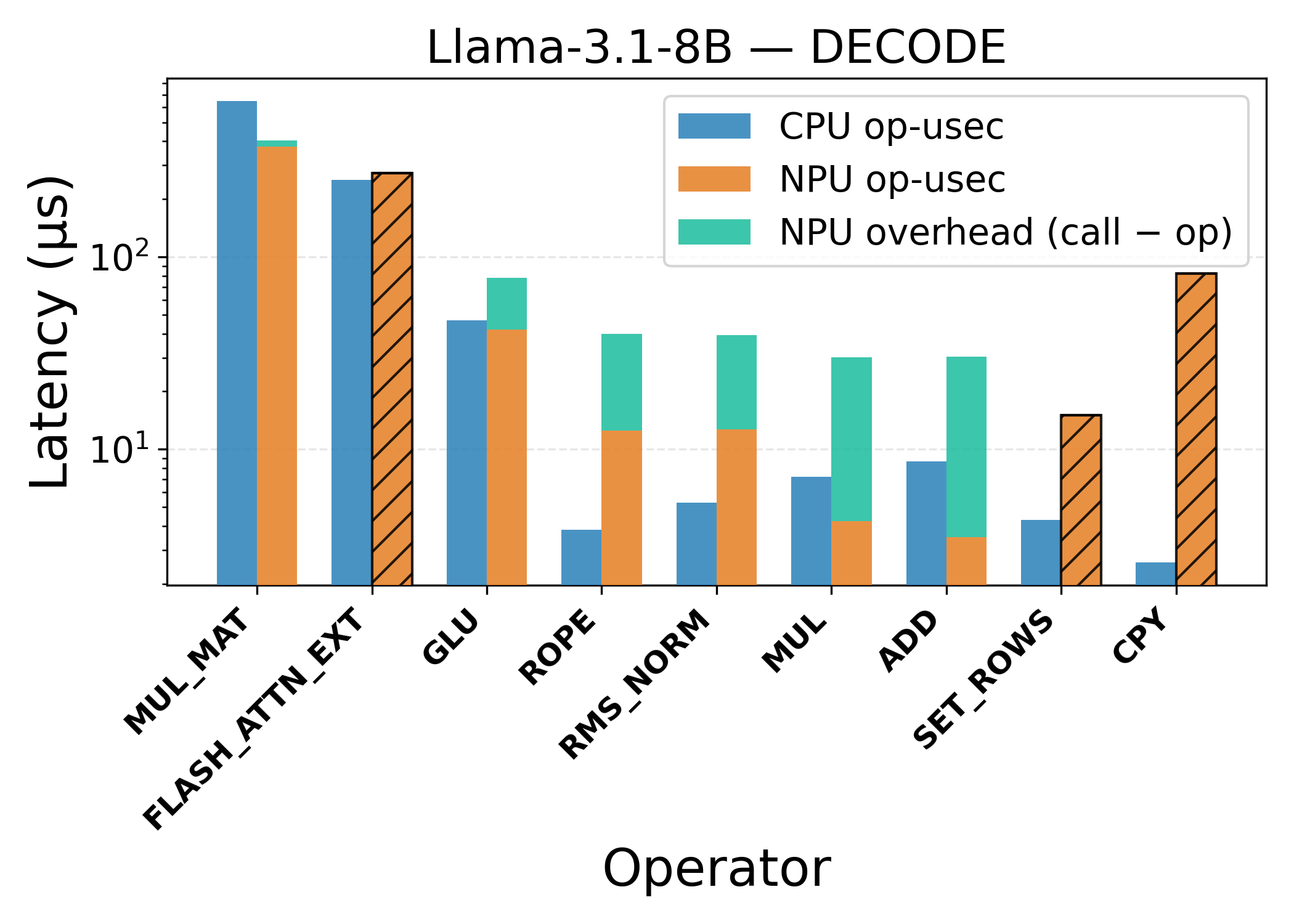}
\centerline{\scriptsize (b) Llama-3.1-8B}
\end{minipage}\hfill

\vspace{0.08cm}

\begin{minipage}[t]{0.48\linewidth}\centering
\includegraphics[width=\linewidth]{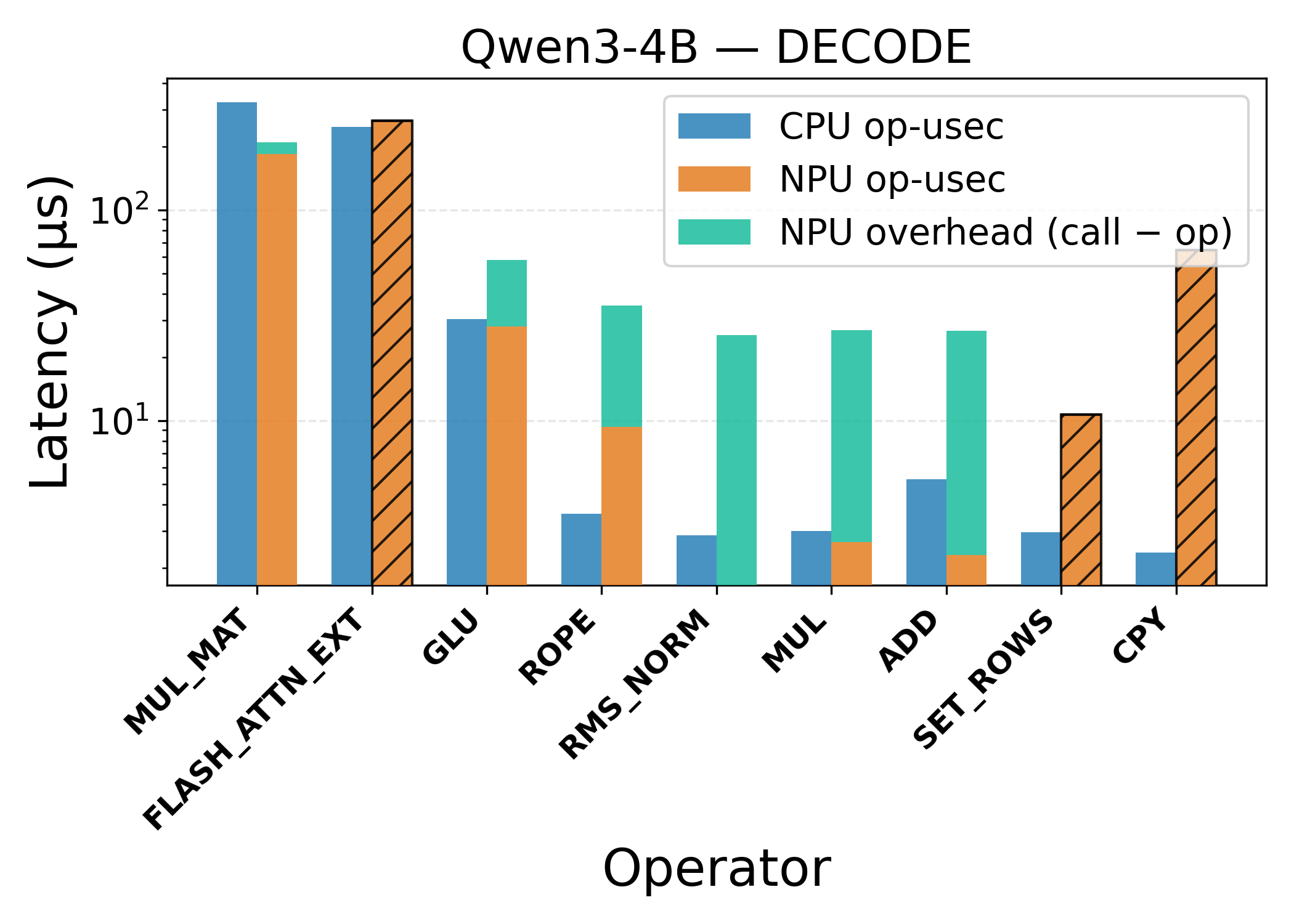}
\centerline{\scriptsize (c) Qwen3-4B}
\end{minipage}\hfill
\begin{minipage}[t]{0.48\linewidth}\centering
\includegraphics[width=\linewidth]{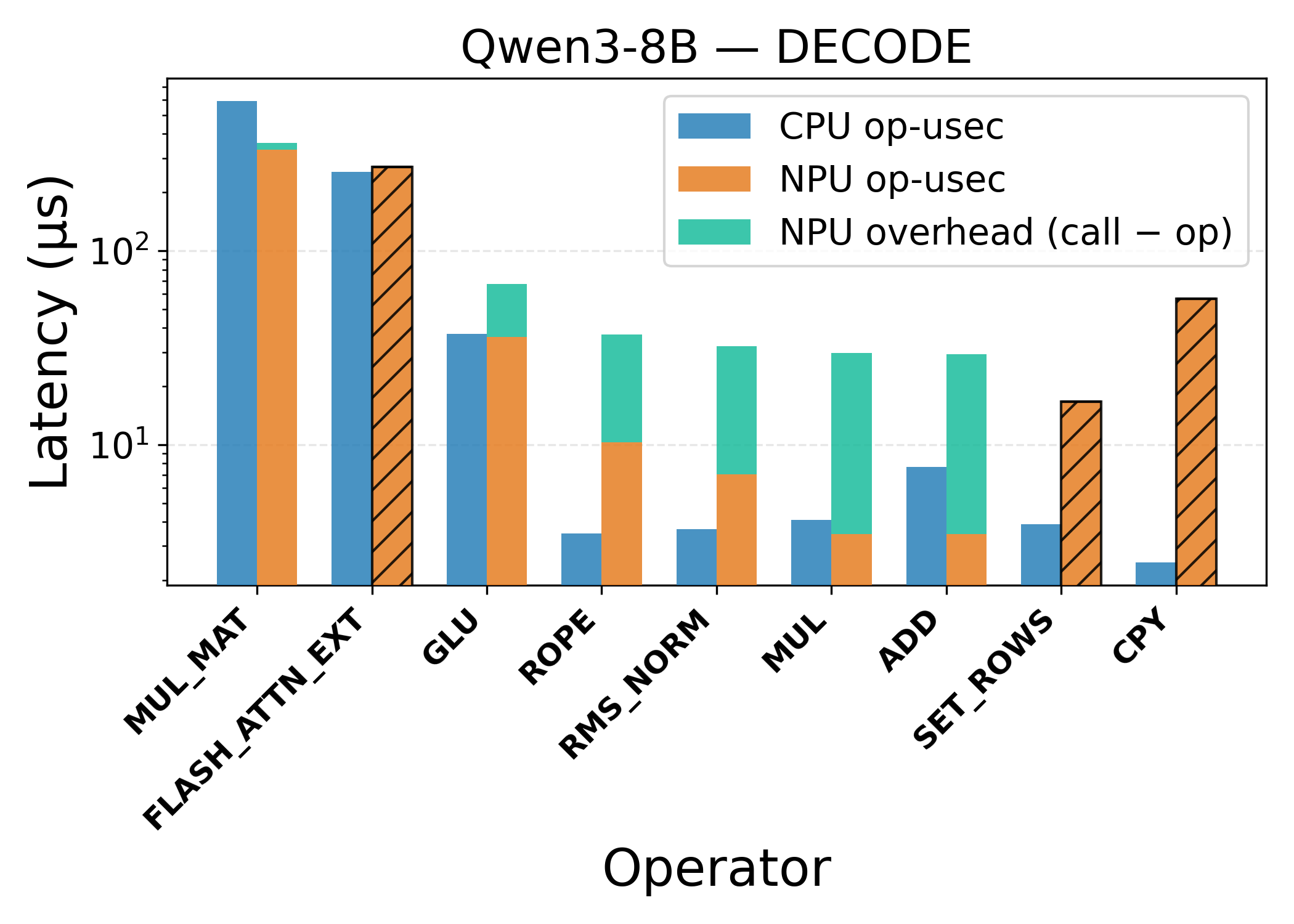}
\centerline{\scriptsize (d) Qwen3-8B}
\end{minipage}

\caption{Decode-stage per-operator latency comparison (op-usec vs.\ call-usec) for CPU and NPU backends. Lightweight operators on the NPU exhibit substantial scheduling overhead.}
\label{fig:decode_overhead}
\end{figure}

\textbf{Scheduling tax on lightweight operators.}
Lightweight operators such as \texttt{RMS\_NORM} and \texttt{ADD} have microsecond-scale computation times on the NPU, but each invocation pays the full cost of CPU--NPU round-trip communication. Their call-usec reaches 8$\times$--22$\times$ their op-usec. Since the Decode stage dispatches hundreds of such operators per token, this ``scheduling tax'' accumulates into a substantial bottleneck.
Fig.~\ref{fig:decode_overhead} visualizes the per-operator latency discrepancy during Decode, showing the dramatic gap between op-usec (pure computation) and call-usec (total invocation) for lightweight operators dispatched to the NPU.

\textbf{Cross-backend fallback penalty.}
The attention operator (\texttt{FLASH\_ATTN\_EXT}) lacks native NPU support and falls back to CPU execution. This fallback incurs 1--1.5$\times$ higher latency than native CPU execution, due to cross-backend cache coherency synchronization, tensor reordering, and memory copy (\texttt{CPY}) overhead. The combined effect of scheduling tax and fallback penalty substantially erodes the NPU's core acceleration advantage.

\subsection{Energy Implications}

Power consumption experiments on Llama-3.2-3B reveal that increasing NPU offloading does \emph{not} reduce total energy consumption. Compared to CPU-only execution, full NPU offloading increases device-level battery drain by 22\% ($\sim$321 token prompt), 32\% ($\sim$ 641 tokens), and 51\% ($\sim$1\,281 tokens). This counter-intuitive result arises from the combined effects of cross-backend scheduling overhead, operator fallback penalties, and increased total execution time under the current heterogeneous path with incomplete operator coverage. Longer prompts amplify this energy penalty, as the Prefill stage, where the NPU is slower, consumes proportionally more time and power.

\subsection{Design Guidelines}
\label{subsec:guidelines}

We distill three guidelines from our quantitative findings:

\textbf{G1: Adopt stage-aware backend selection.}
Route Prefill to multi-core CPUs (1.27--1.62$\times$ faster due to mature GEMM libraries and larger caches) and Decode to the NPU, where GEMV benefits from the NPU's low-latency datapath.

\textbf{G2: Reduce NPU dispatch latency below 10\,$\mu$s.}
Lightweight operators show 8--22$\times$ call-usec/op-usec ratios. Batch dispatch, persistent command queues, or operator fusion are needed to realize the NPU's core advantage.

\textbf{G3: Improve operator coverage.}
Operators used for LLM inference (such as FLASH\_ATTN) should be implemented on the NPU as soon as possible to eliminate the fallback penalty and enhance the NPU's advantages in the decoding phase.

\section{Conclusion}
This paper presents the first stage-aware, operator-level benchmarking study of mobile LLM inference on a CPU–NPU heterogeneous Snapdragon 8 Gen 3. 
We reveal a clear performance divergence: CPUs outperform NPUs in the compute-intensive Prefill stage, while NPUs provide only limited end-to-end gains in Decode. We show that scheduling overhead and cross-backend fallback reduce the practical benefits of NPU offloading. These findings demonstrate the need for stage-aware execution and improved system–hardware co-design to realize efficient on-device LLM inference.


\bibliographystyle{IEEEtran}
\bibliography{ref}

\end{document}